\documentclass[9pt, conference]{IEEEtran}

\usepackage[preprint]{waspaa25}

\usepackage{bm} 
\usepackage{subcaption}

\usepackage {tikz}
\usetikzlibrary{shapes,arrows,positioning,calc,matrix, positioning, intersections, arrows, arrows.meta}
\usetikzlibrary{external}

\title{An Investigation on Combining Geometry and Consistency Constraints into Phase Estimation for Speech Enhancement}

\name{Chun-Wei Ho$^{{1}^{*}}$
      Pin-Jui Ku$^{{1}^{*}}$\thanks{*equal contribution},
      Hao Yen$^{1}$,
      Sabato Marco Siniscalchi$^{1,2}$,
      Yu Tsao$^{1,3}$,
      Chin-Hui Lee$^{2}$}
\address{$^{1}$Georgia Institute of Technology, USA \;
$^{2}$Università degli Studi di Palermo, Italy\\
$^{3}$Academia Sinica, Taiwan
}

\begin{document}

\maketitle

\begin{abstract}
We propose a novel iterative phase estimation framework, termed multi-source Griffin-Lim algorithm (MSGLA), for speech enhancement (SE) under additive noise conditions. The core idea is to leverage the ad-hoc consistency constraint of complex-valued short-time Fourier transform (STFT) spectrograms to address the sign ambiguity challenge commonly encountered in geometry-based phase estimation. Furthermore, we introduce a variant of the geometric constraint framework based on the law of sines and cosines, formulating a new phase reconstruction algorithm using noise phase estimates. We first validate the proposed technique through a series of oracle experiments, demonstrating its effectiveness under ideal conditions. We then evaluate its performance on the VB-DMD and WSJ0-CHiME3 data sets, and show that the proposed MSGLA variants match well or slightly outperform existing algorithms, including direct phase estimation and DNN-based sign prediction, especially in terms of background noise suppression.
\end{abstract}

\section{Introduction}\label{sec:introduction}

Speech enhancement (SE) aims at improving speech quality and intelligibility by suppressing interfering additive noise. While early approaches primarily concentrated on estimating the magnitude of the target spectrogram~\cite{Berouti1979, Lim1979Wiener, Paliwal1987, hu2001speech, srinivasan2006binary, xu2015regression, du2016regression, Wu2017reverberation, qi2020mean, Siniscalchi2021vec2vec}, recent research has increasingly emphasized the importance of accurate phase estimation~\cite{paliwal2011importance, Mowlaee2013Phase, Gerkmann2015Phase} in improving perceptual quality. In response, a variety of model architectures and phase loss functions have been proposed to enhance phase estimation~\cite{erdogan2015phase, takamichi2018phase, liu2019divide, tan2019learning, Tu2020multitarget, ku2023s4nd,lu2023mp, Wang2023tfgridnet, ku2025explicit}. However, estimating phase accurately in noisy environments remains a challenging task~\cite{Alsteris2004importance, Kulmer2015phase, Masuyama2020phase, zhang2024unrestricted}.

A notable strategy in phase estimation for speech enhancement and separation is to leverage geometric constraints~\cite{Mowlaee2012Phase, Wang2019Deep, Mowlaee2013Iterative, Mowlaee2014Time}. These methods assume that the absolute phase difference between sources in a mixture can be determined using the law of cosines~\cite{stewart2015precalculus}, provided that the magnitudes of the sources are known. This reduces the problem of phase estimation to a binary classification task: once the absolute phase difference is computed, the clean speech phase can be obtained by adding or subtracting this difference from the noisy phase, resulting in two possible phase candidates. Nevertheless, one major challenge in this framework lies in resolving the sign ambiguity of the phase difference, especially given the unstructured nature of phase spectrograms. To mitigate this, prior work has incorporated signal processing constraints such as group delay and instantaneous frequency, particularly in harmonic regions~\cite{Mowlaee2012Phase, Mowlaee2014Time}. More recently,~\cite{Wang2019Deep} proposed a sign predictor based on deep neural network (DNN), trained to select the correct sign. However, this approach suffers from the inherent randomness and low predictability of the sign targets~\cite{Wang2021Leveraging}, which can potentially result in an even larger phase deviation when the sign is misclassified.

In this work, we propose a novel framework that addresses the phase reconstruction problem by exploiting the consistency properties of speech and noise signals~\cite{Griffin1984Signal, le2008explicit, ku2025explicit}. Our approach iteratively refines phase estimates using a method similar to the Griffin-Lim algorithm~\cite{Griffin1984Signal}, applied alternately to the speech and noise signals. We refer to this method as the noise magnitude-based multi-source Griffin-Lim algorithm (NM-MSGLA). This iterative process enables the phase estimates to converge toward one of the valid candidates under the geometric constraint. Beyond the conventional geometric formulation based on the noise magnitude, we further introduce a new variant, noise phase-based MSGLA (NP-MSGLA), which leverages noise phase estimates to reconstruct the speech phase using the law of sines and cosines~\cite{stewart2015precalculus}, a formulation not previously explored. This is motivated by our empirical findings that noise phase is often easier to estimate in low-energy regions where clean speech phase is unreliable, suggesting that noise phase can serve as a complementary cue.

We first validate the feasibility of both NM- and NP-MSGLA variants through oracle experiments, where ground-truth speech or noise spectrogram information are available for phase updates. Results show that when either noise magnitude or noise phase is accessible, MSGLA can accurately reconstruct the phase, closely matching the ground truth. After establishing the theoretical performance limit, we evaluate the method in real-world SE tasks using estimated spectrograms on the VB-DMD and WSJ0-CHiME3 data sets. Our results demonstrate that MSGLA consistently matches or outperforms direct phase estimation and shows superior performances compared with the DNN-based sign predictor~\cite{Wang2019Deep}, particularly in background suppression metrics. These findings underscore the effectiveness of MSGLA in leveraging geometry insights and consistency constraints, indicating its strong potential as an effective framework for accurate phase estimation in speech enhancement applications with additive noise.

\section{Related Techniques}

Let \( \mathbf{H}_Y, \mathbf{H}_X, \mathbf{H}_Z \in \mathbb{C}^{M \times N} \) denote the complex-valued short-time Fourier transform (STFT) spectrograms of noisy speech, clean speech, and additive noise, respectively, where \( M \) and \( N \) represent the number of time frames and frequency bins. Under the additive noise assumption, the noisy spectrogram satisfies
\(
\mathbf{H}_Y = \mathbf{H}_X + \mathbf{H}_Z.
\)
The corresponding magnitude and phase spectrograms of noisy speech are:
\[
\mathbf{A}_Y = |\mathbf{H}_Y| \in \mathbb{R}^{M \times N}, \quad \mathbf{P}_Y = \angle \mathbf{H}_Y \in [-\pi, \pi)^{M \times N}.
\]
Similar notations apply to the clean and noise spectrograms.

\subsection{Griffin-Lim Algorithm (GLA)}
\label{subsec:GLA}

The Griffin-Lim Algorithm (GLA)~\cite{Griffin1984Signal} is a classical and widely used method for phase reconstruction. It iteratively alternates between the time and STFT domains, preserving the target magnitude while refining the phase to enforce consistency. Given an \( n \)-th phase estimate \( \mathbf{P}_X^{(n)} \) and the magnitude \( \mathbf{A}_X \) of clean speech, GLA updates it as:
\[
\mathbf{P}_X^{(n+1)} = \angle \left( \text{STFT} \circ \text{iSTFT} \left( \mathbf{A}_X e^{j \mathbf{P}_X^{(n)}} \right) \right),
\]
where \(\text{STFT}\) and \(\text{iSTFT}\) denote the short-time Fourier transform and its inverse, respectively.

\subsection{Phase Estimation with Geometric Constraints}
\label{subsec:phase-estimation-with-geometric}

Algorithms based on geometry constraints utilize trigonometric identities to reconstruct phase under the additive noise assumption. The key idea is that the absolute phase difference between sources in a mixture can be accurately calculated if their magnitudes are known~\cite{Mowlaee2012Phase, Wang2019Deep, Mowlaee2013Iterative, Mowlaee2014Time}. Specifically, for each time-frequency (T-F) bin indexed by \( (m,n) \), the phase difference between the clean and noisy spectrograms could be computed according to the law of cosines~\cite{stewart2015precalculus}:
\begin{align}
    \Delta \mathbf{P}_{m,n} &= (\mathbf{P}_X)_{m,n} - (\mathbf{P}_Y)_{m,n} \\
    &= \pm \cos^{-1}\left(\frac{(\mathbf{A}_Y)_{m,n}^2 + (\mathbf{A}_X)_{m,n}^2 - (\mathbf{A}_Z)_{m,n}^2}{2(\mathbf{A}_X)_{m,n} (\mathbf{A}_Y)_{m,n}}\right),
\end{align}
where \( m, n \) index the time and frequency bins. For simplicity, we omit \( (m,n) \) in the remainder of this paper.

Based on this formulation, a two-stage SE framework~\cite{Mowlaee2012Phase} was proposed. In the first stage, a magnitude estimator predicts the magnitudes of both the speech and noise components, denoted as \( \hat{A}_X \) and \( \hat{A}_Z \). Using these estimates, the absolute value of the phase difference \( |\Delta \hat{\mathbf{P}}| \) and the clean speech phase estimate \( \hat{\mathbf{P}}_X \) are computed as:
\begin{gather}
    |\Delta \hat{\mathbf{P}}| = \cos^{-1}\left(\frac{A_Y^2 + \hat{A}_X^2 - \hat{A}_Z^2}{2\hat{A}_X A_Y}\right) \\
    \label{eq:gc_candidates}
    \hat{\mathbf{P}}_X = \mathbf{P}_Y + \text{sign} \cdot |\Delta \hat{\mathbf{P}}|
\end{gather}
where \( \text{sign} \in \{-1, 1\} \) is determined by a second-stage sign estimator for each entry.
Although this approach effectively transforms the phase estimation problem from a regression task over \( [-\pi, \pi) \) into a binary classification problem, a key challenge lies in accurately determining the sign. The highly unstructured nature of phase spectrograms makes sign estimation difficult. Although signal-based constraints~\cite{Mowlaee2012Phase, Mowlaee2014Time} and DNN-based predictors~\cite{Wang2019Deep} have been proposed, those solutions often struggle with consistency and accuracy due to the random distribution of sign targets~\cite{Wang2021Leveraging}. Incorrect sign predictions can cause large phase deviations, ultimately degrading the quality and intelligibility of reconstructed speech.

\section{Multi-Source Griffin-Lim Algorithm (MSGLA)}

As discussed in Section~\ref{subsec:phase-estimation-with-geometric}, a major limitation of phase estimation based on geometry constraints lies in the high variability and unpredictability of the phase difference signs. To address this, we propose a novel technique that integrates the GLA framework with geometric constraints to iteratively refine phase estimates while enforcing consistency with the phase difference model. Specifically, the proposed multi-mource Griffin-Lim algorithm (MSGLA) alternates GLA-style updates between the clean and noise spectrograms, as illustrated in Figure~\ref{fig:block-diagram-dual-gl}. We introduce two variants of MSGLA, which rely on noise magnitude or noise phase information, respectively.

\subsection{Noise Magnitude-Based MSGLA (NM-MSGLA)}

\begin{figure}[t!]
    \centering
    \resizebox{0.9\linewidth}{!}{\begin{tikzpicture}
    \node[](noisy input){Noisy Spectrogram $\mathbf{A}_Ye^{j\mathbf{P}_Y}$};
    \node[below=.5 of noisy input, draw, label=above:Noise GLA, text width=.72\columnwidth, text height=1.6cm, dashed](replace-noise){};
    \node[draw, text width=.25\columnwidth, right=1.2 of replace-noise.west](replace-noise-inside) {Replace with \\ $\hat A_z$ or $\hat P_z$};
    \node[draw, right=1.7 of replace-noise-inside, text width=.1\columnwidth, align=center](replace-noise-istft) {iSTFT\\$|$\\STFT};
    \node[above=.5 of noisy input, draw, label=above:Speech GLA, text width=.72\columnwidth, text height=1.4 cm, dashed](replace-clean){};
    \node[draw, left=1.4 of replace-clean.east](replace-clean-inside) {Replace with $\hat A_x$};
    \node[draw, left=1.4 of replace-clean-inside, text width=.1\columnwidth, align=center](replace-clean-istft) {iSTFT\\$|$\\STFT};

    \node[left=1.3 of noisy input](sum-left){$\bigoplus$};
    \node[right=1.3 of noisy input](sum-right){$\bigoplus$};
    \node[above=.5 of replace-clean](clean-feat){Estimated Speech Magnitude $\hat{\mathbf{A}}_X$};
    \node[below=.3 of replace-noise](noise-feat){Estimated Noise Magnitude $\hat{\mathbf{A}}_Z$ or Noise Phase $\hat{\mathbf{P}}_Z$};
    \node[anchor=north, minimum width=8.7cm, text height=2.6cm, fill=red, fill opacity=0.2, text opacity=1, text=red] at (noisy input){};
    \node[anchor=south, minimum width=8.7cm, text depth=2.8cm, fill=green, fill opacity=0.2, text opacity=1, text=teal] at (noisy input){};

    \draw[->] (replace-noise.west) --node[above]{$A_ze^{jP_z}$} (replace-noise-inside);
    \draw[->] (replace-noise-inside) --node[above]{$\hat A_ze^{jP_z}$} node[below]{or $A_ze^{j \hat P_z}$} (replace-noise-istft);
    \draw[-] (replace-noise-istft) -- (replace-noise.east);

    \draw[->] (replace-clean.east) --node[above]{$A_Xe^{jP_X}$} (replace-clean-inside);
    \draw[->] (replace-clean-inside) --node[above]{$\hat A_X e^{jP_X}$} (replace-clean-istft);
    \draw[-] (replace-clean-istft) -- (replace-clean.west);

    \draw[->] (noisy input) --node[above, pos=0.8]{$+$} (sum-left);
    \draw[->] (noisy input) --node[below, pos=0.8]{$+$} (sum-right);
    \draw[->] (replace-clean) -|node[left, pos=0.85]{$-$} (sum-left);
    \draw[->] (replace-noise) -|node[right, pos=0.85]{$-$} (sum-right);
    \draw[-] (sum-left) |- (replace-noise);
    \draw[-] (sum-right) |- (replace-clean);
    \draw[->] (clean-feat.south-|replace-clean-inside.30) -- (replace-clean-inside.30);
    \draw[->] (noise-feat.north-|replace-noise-inside) -- (replace-noise-inside);
\end{tikzpicture}}
    \caption{A block diagram of the proposed MSGLA. The iSTFT-STFT block represents the operation of $\text{STFT} \circ \text{iSTFT}(\cdot)$.}
    \label{fig:block-diagram-dual-gl}
\end{figure}

Given the estimated magnitude spectrograms \( \hat{\mathbf{A}}_X \) and \( \hat{\mathbf{A}}_Z \), and the \( n \)-th iteration of the clean speech phase estimate \( \hat{\mathbf{P}}_X^{(n)} \), NM-MSGLA updates the phase as follows:
\begin{align}
    \tilde{\mathbf{P}}_X^{(n+1)} &= \angle \left( \text{STFT} \circ \text{iSTFT} \left( \hat{\mathbf{A}}_X e^{j\hat{\mathbf{P}}_X^{(n)}} \right) \right) \\
    \tilde{\mathbf{P}}_Z^{(n+1)} &= \angle \left( \text{STFT} \circ \text{iSTFT} \left( \mathbf{H}_Y - \hat{\mathbf{A}}_X e^{j\tilde{\mathbf{P}}_X^{(n+1)}} \right) \right) \\
    \hat{\mathbf{P}}_X^{(n+1)} &= \angle \left( \mathbf{H}_Y - \hat{\mathbf{A}}_Z e^{j\tilde{\mathbf{P}}_Z^{(n+1)}} \right)
\end{align}

In essence, NM-MSGLA enforces both consistency (via GLA-like projections) and geometric constraints (via the additive noise model). Empirically, we observe that after a few iterations (typically around 5), the estimated phase \( \hat{\mathbf{P}}_X^{(n)} \) converges to either \( \mathbf{P}_Y + |\Delta \hat{\mathbf{P}}| \) or \( \mathbf{P}_Y - |\Delta \hat{\mathbf{P}}| \), thereby implicitly resolving the sign ambiguity in Equation~\eqref{eq:gc_candidates}.

\subsection{Noise Phase-Based MSGLA (NP-MSGLA)}

In Section~\ref{subsec:phase-estimation-with-geometric}, we described how two phase candidates for clean speech can be derived using the law of cosines when both speech and noise magnitudes are available. Interestingly, a complementary formulation can be obtained using the law of sines~\cite{stewart2015precalculus} when the speech magnitude and noise phase are known. Specifically, given the estimated speech magnitude \( \hat{\mathbf{A}}_X \) and the noise phase \( \hat{\mathbf{P}}_Z \), the clean speech phase \( \hat{\mathbf{P}}_X \) should satisfy:
\begin{align}
\label{eq:sine_law}
    \frac{\mathbf{A}_Y}{\sin(\hat{\mathbf{P}}_X - \hat{\mathbf{P}}_Z)} = \frac{\hat{\mathbf{A}}_X}{\sin(\mathbf{P}_Y - \hat{\mathbf{P}}_Z)}
\end{align}
Rearranging Equation~\eqref{eq:sine_law} yields two valid solutions for \( \hat{\mathbf{P}}_X \):
\begin{align}
\label{eq:noise-pha-candidates}
    \hat{\mathbf{P}}_X = \begin{cases}
        \sin^{-1}\left(\frac{\mathbf{A}_Y}{\hat{\mathbf{A}}_X} \sin(\mathbf{P}_Y - \hat{\mathbf{P}}_Z)\right) + \hat{\mathbf{P}}_Z \\
        \pi - \sin^{-1}\left(\frac{\mathbf{A}_Y}{\hat{\mathbf{A}}_X} \sin(\mathbf{P}_Y - \hat{\mathbf{P}}_Z)\right) + \hat{\mathbf{P}}_Z
    \end{cases}
\end{align}

With this geometric relation, NP-MSGLA updates the phase as:
\begin{align}
    \tilde{\mathbf{P}}_X^{(n+1)} &= \angle \left( \text{STFT} \circ \text{iSTFT} \left( \hat{\mathbf{A}}_X e^{j\hat{\mathbf{P}}_X^{(n)}} \right) \right) \\
    \tilde{\mathbf{A}}_Z^{(n+1)} &= \left| \text{STFT} \circ \text{iSTFT} \left( \mathbf{H}_Y - \hat{\mathbf{A}}_X e^{j\tilde{\mathbf{P}}_X^{(n+1)}} \right) \right| \\
    \hat{\mathbf{P}}_X^{(n+1)} &= \angle \left( \mathbf{H}_Y - \tilde{\mathbf{A}}_Z^{(n+1)} e^{j\hat{\mathbf{P}}_Z} \right)
\end{align}

Again, we find that NP-MSGLA converges to one of the two phase candidates derived in Equation~\eqref{eq:noise-pha-candidates}, thereby resolving the phase ambiguity in an unsupervised manner.

\begin{figure}[t]
    \centering
    \begin{subfigure}{0.9\linewidth}
        \centering
        \includegraphics[width=\linewidth]{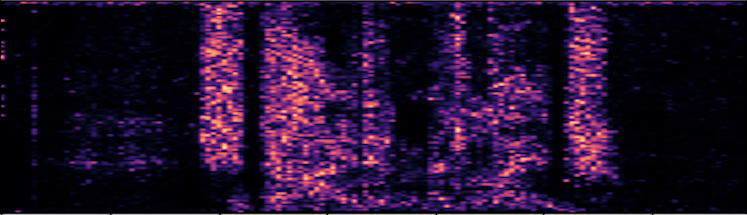}
        \caption{Noise phase estimation error}
        \label{fig:subfig1}
    \end{subfigure}
    \begin{subfigure}{0.9\linewidth}
        \centering
        \includegraphics[width=\linewidth]{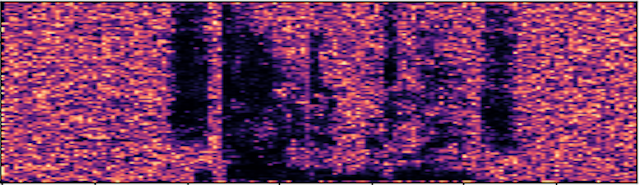}
        \caption{Clean speech phase estimation error}
        \label{fig:subfig2}
    \end{subfigure}
    \vskip 1em
    \begin{subfigure}{0.9\linewidth}
        \centering
        \includegraphics[width=\linewidth]{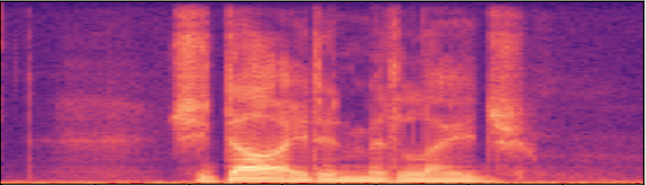}
        \caption{Clean speech log-magnitude spectrogram}
        \label{fig:subfig3}
    \end{subfigure}
    \caption{(a)(b) Comparison of noise and speech phase spectrogram estimation errors. Dark regions indicate low or near-zero estimation error. (c) The corresponding magnitude spectrogram in log scale.}
    \label{fig:noise_speech_phase_error_comparison}
\end{figure}

While the geometric constraints phase estimation approach with two phase candidates based on the law of cosines have been studied for over a decade~\cite{Mowlaee2012Phase,Mowlaee2013Iterative}, to the best of our knowledge, this work is the first to formulate phase candidate selection using the law of sines. This direction has remained unexplored due to the perceived difficulty of estimating the noise phase spectrogram, an objective often considered as challenging as estimating the speech phase itself. However, we argue that it is worth exploring, motivated by some empirical observations from a DNN-based direct phase estimation model proposed in~\cite{lu2023mp}. As shown in Fig.~\ref{fig:noise_speech_phase_error_comparison}, clean speech phase tends to be accurately estimated in regions with strong speech energy, whereas noise phase is easier to estimate in low-energy regions. This complementary pattern suggests that noise phase can serve as a valuable cue where clean phase estimation fails. Our proposed NP-MSGLA is the first to exploit this insight, using the law of sines to derive valid phase candidates from the estimated speech magnitude and noise phase, offering a principled framework for phase reconstruction under challenging conditions.

\section{Experimental Setup}

\subsection{Data Sets}
Two well-known SE benchmark data sets, the VoiceBank-DEMAND (VB-DMD)~\cite{ValentiniBotinhao2017} and WSJ0-CHiME3~\cite{lemercier2023storm}, are adopted for our experiments. For the VB-DMD data set, we followed the commonly used training, validation, and testing splits as detailed in \cite{Lu2022conditional}, with speakers "p286" and "p287" designated for the validation set. This setup provided us with 10,802 training utterances, 770 validation utterances, and 824 test utterances. The WSJ0-CHiME3 data set was prepared according to the instructions in~\cite{lemercier2023storm}, resulting in 12,776 training utterances, 1,206 validation utterances, and 651 test utterances. The \textit{Remix} and \textit{BandMask} data augmentation methods proposed in \cite{Dfossez2020real} are adopted. Our evaluation metrics include wide-band perceptual evaluation of speech quality (PESQ)~\cite{rix2001pesq}, extended short-term objective intelligibility (ESTOI)~\cite{Jensen2016stoi}, scale-invariant signal-to-noise ratio (SI-SNR)~\cite{Roux2019sisdr}. and prediction of rating of background distortion (CBAK)~\cite{Hu2008evaluation}.

\subsection{Model Architecture}
\begin{figure}
    \centering
    \includegraphics[width=\linewidth]{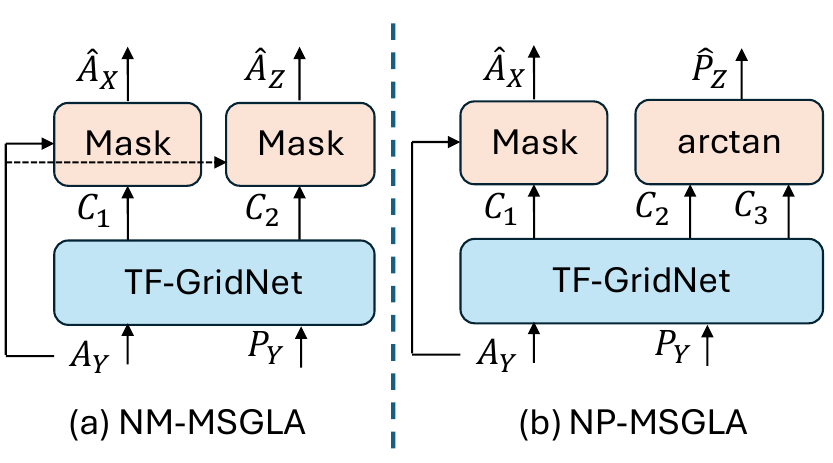}
    \caption{Illustration of inputs and outputs for the two MSGLA variants.}
    \label{fig:illustration}
\end{figure}
We use TF-GridNet~\cite{Wang2023tfgridnet} as the backbone model for all experiments, adopting the architecture and hyperparameters from~\cite{Wang2023tfgridnet} with \( D=64 \), \( H=128 \), \( I=J=1 \), and \( B=3 \). Note that we exclude the global attention layer as we discovered limited performance improvement, resulting in approximately 1.3 million trainable parameters.

Fig.~\ref{fig:illustration} illustrates the inputs and outputs adopted in TF-GridNet in our experiments. For NM-MSGLA, the model takes the noisy magnitude and phase spectrograms \( (\mathbf{A}_{Y}, \mathbf{P}_{Y}) \) as input with two output channels $\mathbf{C}_1, \mathbf{C}_2 \in [0, 1]^{M \times N}$: one used as a real-valued mask to estimate the speech magnitude \( \hat{\mathbf{A}}_X \), and the other to estimate the noise magnitude \( \hat{\mathbf{A}}_Z \). For NP-MSGLA, the model produces three output channels $\mathbf{C}_1, \in [0, 1]^{M \times N}$  and $\mathbf{C}_2, \mathbf{C}_3 \in \mathbb{R}^{M \times N}$: one for estimating \( \hat{\mathbf{A}}_X \), and two channels \( \mathbf{C}_2, \mathbf{C}_3 \in \mathbb{R}^{M \times N} \) used to compute the estimated noise phase via
\(
\hat{\mathbf{P}}_Z = \arctan \left( \frac{\mathbf{C}_2}{\mathbf{C}_3} \right).
\)
An \( L_1 \) loss is applied to the magnitude predictions, while phase estimation error is computed using a cosine distance loss with phase derivative terms~\cite{takamichi2018phase, Masuyama2020phase} between the original and estimated noise phase spectrograms.

\begin{table}[t]
\caption{Comparison of phase estimation results using either oracle or estimated information for MSGLA using the VB-DMD data set.}
\label{table:oracle_dgla}
\centering
\setlength\tabcolsep{6.0pt}

\begin{subtable}[t]{0.95\linewidth}
    \centering
    \caption{NM-MSGLA: Using speech and noise magnitudes.}
    \label{table:nm_dgla}
    \resizebox{\linewidth}{!}{%
        \begin{tabular}{cccccc}
            \toprule
            $\mathbf{A}_x$ & $\mathbf{A}_z$ & PESQ & ESTOI & SI-SNR & Cos. Sim. \\
            \midrule
            Oracle   & Oracle     & 3.60 & 0.91 & 22.43 & 0.87 \\
            Oracle   & Estimated  & 3.48 & 0.89 & 19.41 & 0.45 \\
            Estimated & Oracle    & 3.45 & 0.90 & 21.02 & 0.74 \\
            Estimated & Estimated & 3.39 & 0.88 & 19.16 & 0.41 \\
            \bottomrule
        \end{tabular}
    }
\end{subtable}

\vspace{1.2em}

\begin{subtable}[t]{0.95\linewidth}
    \centering
    \caption{NP-MSGLA: Using speech magnitude and noise phase.}
    \label{table:np_dgla}
    \resizebox{\linewidth}{!}{%
        \begin{tabular}{cccccc}
            \toprule
            $\mathbf{A}_x$ & $\mathbf{P}_z$ & PESQ & ESTOI & SI-SNR & Cos. Sim. \\
            \midrule
            Oracle   & Oracle     & 3.55 & 0.91 & 21.55 & 0.78 \\
            Oracle   & Estimated  & 3.46 & 0.89 & 19.55 & 0.47 \\
            Estimated & Oracle    & 3.43 & 0.90 & 21.02 & 0.70 \\
            Estimated & Estimated & 3.36 & 0.88 & 19.54 & 0.47 \\
            \bottomrule
        \end{tabular}
    }
\end{subtable}
\end{table}

All models are trained using 32-second input segments over 160 epochs, optimized with Adam~\cite{kingma2017adam}. The initial learning rate is set to \( 1 \times 10^{-3} \) and decayed to \( 1 \times 10^{-5} \) using cosine annealing. STFT is performed using a 512-sample Hann window with a 256-sample hop size. Magnitude compression follows the scheme in~\cite{lemercier2023storm} with parameters \( a = 0.5 \) and \( b = 1 \).

\section{Experimental Results and Analysis}
\subsection{Phase Reconstruction with Oracle Noise Information}

We first evaluate the effectiveness of NM-MSGLA and NP-MSGLA using either oracle or estimated versions of \( \mathbf{A}_X \), \( \mathbf{A}_Z \), and \( \mathbf{P}_Z \) for speech phase reconstruction on the VB-DMD data set. Each method is run for 5 iterations, initialized with the noisy phase \( \mathbf{P}_Y \). The resulting phase estimates are evaluated by their average cosine similarity with the ground truth \( \mathbf{P}_X \), and are also combined with the estimated speech magnitude \( \hat{\mathbf{A}}_X \) for standard speech quality metrics.

Table~\ref{table:nm_dgla} presents the results for NM-MSGLA. As expected, using oracle speech and noise magnitudes yields the best performance, achieving a 0.87 cosine similarity with the ground truth phase. Interestingly, we can observed strong results  when the oracle noise magnitude is used, with an SI-NSR of 21.02, and a cosine similarity equal to 0.74. Moreover, the comparison of the results obtained with only noise oracle magnitude and only speech oracle magnitudes show that noise magnitude plays a more important role in phase reconstruction. This highlights the critical role of accurate noise magnitude in phase reconstruction. Finally, a performance drop is reported when both magnitudes are estimated, with a 0.09 degradation in PESQ compared to the case with oracle speech magnitude.

Similar trends are observed in Table~\ref{table:np_dgla} for NP-MSGL. With oracle speech magnitude and noise phase, the method achieves 3.55 PESQ and 0.78 cosine similarity. Replacing speech magnitude with its estimated counterpart results in minor drops in SI-SNR and similarity, but a more noticeable PESQ decrease to 3.43. When both inputs are estimated, SI-SNR and cosine similarity remain stable, though PESQ drops by 0.1 compared to the oracle speech magnitude case. These oracle experiments confirm that the proposed MSGLA variants—by leveraging both geometric and consistency constraints—provide a viable and effective framework for phase reconstruction.

\subsection{Phase Reconstruction with Estimated Noise}
\begin{table}[t]
\caption{Comparisons of speech enhancement performances on the VB-DMD data set}
\label{table:enh_vbdmd}
\begin{center}
  \setlength\tabcolsep{4.0pt} 
  \resizebox{\columnwidth}{!}{%
    \begin{tabular}{cccccc}
      \toprule
      Algorithm   & MetricGAN & PESQ & ESTOI & SI-SNR & CBAK \\
      \midrule
      Noisy Speech     & --        &  1.97    &  0.72     &   8.40   &   2.55   \\
      \midrule
      phase estimator   & N         &  3.31    &   0.88    &   19.48   &   3.10   \\
      GLA             & N         & 3.37    &  0.88    &  18.65   &  3.09    \\
      sign predictor & N         &  3.37    &   \textbf{0.89}    &   19.41   &  3.15    \\
      NM-MSGLA         & N         & \textbf{3.39}    &  0.88    &  19.16   & 3.14 \\
      NP-MSGLA         & N         & 3.36     & 0.88    &  \textbf{19.54}   & \textbf{3.16}     \\
      \midrule
      phase estimator   & Y      &   3.44   &    \textbf{0.89}    &   19.13   & 3.13     \\
      GLA             & Y         &   3.45   &   0.88    &  18.68    &  3.13    \\
      sign predictor & Y         &  3.41    &    \textbf{0.89}    &   19.16   &   3.11   \\
      NM-MSGLA         & Y         &   3.44   &   0.88    &  19.25    &   3.17  \\
      NP-MSGLA         & Y         &   \textbf{3.46}   &   \textbf{0.89}    &  \textbf{19.61}    &   \textbf{3.18}   \\
      \bottomrule
    \end{tabular}
  }
\end{center}
\end{table}

We now evaluate the proposed MSGLA methods under a blind setting and compare them with three baselines. The first baseline, referred to as the \textbf{direct phase estimator}, shares the same architecture as NP-MSGLA, except that the second and third output channels are directly used to compute the speech phase via
\(
\hat{\mathbf{P}}_X = \arctan \left( \frac{\mathbf{C}_2}{\mathbf{C}_3} \right)
\)
without applying any geometric constraints. The cosine distance loss with phase derivative terms is used to compute the error between the estimated speech phase \(\hat{\mathbf{P}}_X\) and the ground-truth phase \(\mathbf{P}_X\). The second baseline is the conventional \textbf{GLA} described in Section~\ref{subsec:GLA}, where the predicted speech magnitude \(\hat{\mathbf{A}}_X\) from NM-MSGLA is used as input for phase reconstruction, and the initial phase estimate \(\hat{\mathbf{P}}_X^{(0)}\) is set to the noisy phase \(\mathbf{P}_Y\).

The third baseline is the \textbf{DNN-based sign predictor} introduced in~\cite{Wang2019Deep}, which uses geometric constraints to compute the absolute phase difference and then predicts the sign using a DNN-based model. Specifically, we implement this using an additional compact TF-GridNet model with hyperparameters set to \( D=32 \), \( H=64 \), \( I=J=1 \), and \( B=2 \), resulting in approximately 219K additional trainable parameters. This model takes as input the noisy spectrogram \((\mathbf{A}_Y, \mathbf{P}_Y)\), along with the estimated magnitudes \(\hat{\mathbf{A}}_X\) and \(\hat{\mathbf{A}}_Z\) from NP-MSGLA. It outputs a single channel \(\mathbf{C} \in \mathbb{R}^{M \times N}\), which is used to compute the final speech phase estimate as
\[
    \hat{\mathbf{P}}_X = \mathbf{P}_Y + \tanh(\mathbf{C}) \cdot \cos^{-1}\left(\frac{A_Y^2 + \hat{A}_X^2 - \hat{A}_Z^2}{2\hat{A}_X A_Y}\right)
\]
In addition to these baselines, we also investigate the effect of incorporating the MetricGAN loss~\cite{fu2019metricGAN} during training to further improve perceptual quality.

The evaluation results on the VB-DMD and WSJ0-CHiME3 data sets are presented in Tables~\ref{table:enh_vbdmd} and \ref{table:enh_wsj0_chime3}, respectively. Notably, GLA consistently achieves strong PESQ scores, especially on the WSJ0-CHiME3 data set where the SNR value could be low as -6 dB, highlighting the importance of phase consistency in producing perceptually high-quality speech. However, its SI-SNR scores remain relatively low, suggesting that although GLA generates realistic-sounding speech, its estimated phase may deviate more from the ground truth compared to other methods.

Across both sets, the proposed variants, NM-MSGLA and NP-MSGLA,demonstrate competitive, and in some cases marginally improved, performance compared to both the direct phase estimator and the DNN-based sign predictor. In particular, NP-MSGLA achieves the highest SI-SNR and CBAK scores on the VB-DMD data set, indicating better signal reconstruction and reduced background interference.
These results support our central hypothesis: combining geometric constraints with consistency-based refinement provides an effective approach to phase estimation in SE. Our preliminary results are encouraging and the proposed MSGLA framework could open new avenues for robust phase reconstruction in speech enhancement.

\begin{table}[t]
\caption{Comparisons of speech enhancement performances on the WSJ0-CHiME3 data set}
\label{table:enh_wsj0_chime3}
\begin{center}
  \setlength\tabcolsep{4.0pt} 
  \resizebox{\columnwidth}{!}{%
    \begin{tabular}{cccccc}
      \toprule
      Algorithm   & MetricGAN & PESQ & ESTOI & SI-SNR & CBAK \\
      \midrule
      Noisy Speech     & --        & 1.35  & 0.63   &  4.00    &  2.07    \\
      \midrule
      phase estimator  & N         &  2.95    &    \textbf{0.90}   &  \textbf{15.88}    &  2.45    \\
      GLA             & N         & \textbf{3.01}      &  0.88     & 14.01  & \textbf{2.46}  \\
      sign predictor & N         &  2.93    &   0.89    &   15.36   &   2.45   \\
      NM-MSGLA         & N         & 2.95  &    0.89   &  14.89    &  2.45
      \\
      NP-MSGLA         & N         & 2.96  &    0.89   &  15.57    & 2.45     \\
      \midrule
      phase estimator   & Y         &   3.00   &    \textbf{0.90}   &   \textbf{15.67}   &  2.46    \\
      GLA             & Y         &  \textbf{3.10}     &   0.88    &   13.58   &  2.50    \\
      sign predictor & Y         &  3.06    &   \textbf{0.90} &   14.95   &  2.48    \\
      NM-MSGLA         & Y         & 3.09      &  0.89    &  14.77  &  \textbf{2.51} \\
      NP-MSGLA         & Y         & 3.07     &  0.89     & 15.57  &   \textbf{2.51}   \\
      \bottomrule
    \end{tabular}
  }
\end{center}
\end{table}

\section{Conclusion}
In this paper, we propose a multi-source Griffin-Lim algorithm (MSGLA), a novel iterative approach to phase estimation for speech enhancement that unifies geometry and consistency constraints. Unlike prior techniques that rely solely on magnitude-based geometry formulations or direct phase regression, MSGLA iteratively refines phase estimates through alternating GLA-like updates. Furthermore, we introduce NM-MSGLA and NP-MSGLA, which, for the first time, leverages noise phase estimates and the law of sines and cosines to derive valid speech phase candidates. Experimental results under both oracle and blind settings validate the effectiveness of our proposed techniques, showing competitive or superior performance compared to direct phase estimation and DNN-based sign prediction.


\clearpage
\bibliographystyle{IEEEtran}
\bibliography{references/refs}

\end{document}